\documentclass[%
reprint,
amsmath,amssymb,
aps,
]{revtex4-1}

\usepackage{rotating}
\usepackage{color}
\usepackage{algorithm}
\usepackage[noend]{algpseudocode}
\usepackage{graphicx}% Include figure files
\usepackage{dcolumn}% Align table columns on decimal point
\usepackage{bm}% bold math
\usepackage{hyperref}
\usepackage{array}
\makeatletter
\newcommand*{\algrule}[1][\algorithmicindent]{\makebox[#1][l]{\hspace*{.5em}\thealgruleextra\vrule height \thealgruleheight depth \thealgruledepth}}%
\newcommand*{\thealgruleextra}{}
\newcommand*{\thealgruleheight}{.75\baselineskip}
\newcommand*{\thealgruledepth}{.25\baselineskip}

\newcount\ALG@printindent@tempcnta
\def\ALG@printindent{%
	\ifnum \theALG@nested>0% is there anything to print
	\ifx\ALG@text\ALG@x@notext% is this an end group without any text?
	\else
	\unskip
	\addvspace{-1pt}% FUDGE to make the rules line up
	\ALG@printindent@tempcnta=1
	\loop
	\algrule[\csname ALG@ind@\the\ALG@printindent@tempcnta\endcsname]%
	\advance \ALG@printindent@tempcnta 1
	\ifnum \ALG@printindent@tempcnta<\numexpr\theALG@nested+1\relax% can't do <=, so add one to RHS and use < instead
	\repeat
	\fi
	\fi
}%
\usepackage{etoolbox}
\patchcmd{\ALG@doentity}{\noindent\hskip\ALG@tlm}{\ALG@printindent}{}{\errmessage{failed to patch}}
\makeatother

\newbox\statebox
\newcommand{\myState}[1]{%
	\setbox\statebox=\vbox{#1}%
	\edef\thealgruleheight{\dimexpr \the\ht\statebox+1pt\relax}%
	\edef\thealgruledepth{\dimexpr \the\dp\statebox+1pt\relax}%
	\ifdim\thealgruleheight<.75\baselineskip
	\def\thealgruleheight{\dimexpr .75\baselineskip+1pt\relax}%
	\fi
	\ifdim\thealgruledepth<.25\baselineskip
	\def\thealgruledepth{\dimexpr .25\baselineskip+1pt\relax}%
	\fi
	\State #1%
	\def\thealgruleheight{\dimexpr .75\baselineskip+1pt\relax}%
	\def\thealgruledepth{\dimexpr .25\baselineskip+1pt\relax}%
}
\newcommand\Tstrut{\rule{0pt}{2.6ex}}
\newcommand\Bstrut{\rule[-0.9ex]{0pt}{0pt}}

\newcolumntype{L}{>{\centering\arraybackslash}m{3cm}}

\begin{document}
	
\title{A Fast, Parallel Algorithm for Distant-dependent Calculation of Crystal Properties}
%\title{High Precision Calculations of the Lennard-Jones Lattice Constants in Five Lattices}
\author{Matthew Stein}
 \email{mstein@smu.edu}
\affiliation{%
 	Southern Methodist University\\
 	3215 Daniel Ave. \\
 	Dallas, Texas, 75205, USA
 }%

\date{\today}

\begin{abstract}
	
	A fast, parallel algorithm for distant-dependent calculation and simulation of crystal properties is presented along with speedup results and methods of application.  An illustrative example is used to compute the Lennard-Jones lattice constants up to 32 significant figures for $4\leq p\leq30$ in the simple cubic, face-centered cubic, body-centered cubic, hexagonal-close-pack, and diamond lattices. In most cases, the known precision of these constants is more than doubled, and in some cases, corrected from previously published figures. The tools and strategies to make this computation possible are detailed along with application to other potentials, including those that model defects.
	
	%The lattice constants, $L_p$, useful for computing the Lennard-Jones potential, have been extended up to 32 significant figures, and in some cases, corrected from previous publication.  The $L_p$ terms are given for $4\leq p\leq30$ in the simple cubic, face-centered cubic, body-centered cubic, hexagonal-close-pack (HCP), and diamond (DIA) lattices. This precision was obtained through the use of fast, parallel algorithms which exploit the symmetry of each lattice, including a novel approach for the DIA and HCP lattices. The tools and strategies to make this computation possible are detailed.\\
	
	
\end{abstract}

\maketitle

\section{Introduction}

Calculations of crystal potentials or force interactions, whether through molecular dynamics or classical potentials, will rely on functions of distances between many atoms.  In either case, computational complexity and time will limit the precision with which values are calculated.  Even in the case of classical potentials, which are less computationally intense, crystal simulations and calculations are usually limited to the millions of atoms, with determined values often having fewer significant figures than a single-precision float.

Classical potential fitting has also become more complex in attempts to adapt a single model to a greater number of situations.  The Lennard-Jones potential \cite{Jones636} is simple and widely used for its computational speed, but much more accurate models exist.  The Buckingham potential \cite{Buckingham1938} expanded on the Lennard-Jones potential, replacing the Pauli repulsive term with an exponential function but at computational cost.  The Stillinger-–Weber potential \cite{SW1985} (hereafter SW potential) was proposed as a further improvement, now taking into account not just distance between atoms but also the angles of their bonds in a new 3-body term.  

Improvements on the classical potentials have thus progressed for decades \cite{SWex1,SWex2,SWex3,Pizza2013}, with attempts to find a potential model that works not only with perfects crystals, but those with point defects, plane defects, and more.  A fitted formula in one situation (temperature, lattice, atomic composition) often does not suitably agree with experimental values from another.  As such, the potentials grow ever more complex, and determining parameters comes at greater cost, but the objective of a transferable model remains a priority.

Rather than limiting calculations to a small number of atoms (and thus limited precision), or expanding compute time (which schedules and resources may not permit), a faster optimized algorithm could be used to achieve better and/or less costly results.
%Any separable constant terms in summations would be more quickly determined, and at higher precisions.
Additionally, potentials with arbitrary cut-off values (often used to shorten compute time) can be relaxed for better fitting of other parameters and more realistic simulation.  An adaptive algorithm would also ideally be suited for studies of non-ideal lattices with defects, vacancies or other imperfections.

The inclusion of contributions from further atoms or those with defect locations should also come with questions about the precision of the calculation.  For example, a single interstitial sufficiently far away from a reference atom may not affect the total potential energy, but a plane defect at the same distance may have significant contributions when all atoms across the plane are considered.  It may be useful to use very high-precision variables in computation, further advancing the need for a faster algorithm.

%With the increasing use of double-, long-double-, and quadruple-precision floats (64-, 80-, and 128-bit respectively) displacing lower-precision types, and with new hardware being designed to handle 128-bit types natively, it is useful to extend and correct the precision and values of the $L_p$ terms.
%%for use with experiment and simulation.  
%The primary obstacles to achieving higher precision have been computational power and a lack of support for high-precision floats, so with access to cluster resources and C++ compatible packages such as the Portable, Extensible Toolkit for Scientific Computation (PETSc) (which contains both 128-bit floats and the Message Passing Interface (MPI) communicators to handle such variables) it is now feasible to overcome these challenges.  Even with these resources, however, the incredible number of calculations necessary to converge (\ref{eq:lp}) to high precision still requires a new algorithmic approach to provide significant speedup.

\section{Computational Approach}

Potential and force calculations in a crystal depend on distances between pairs of atoms.  Any summation over lattice points will first require the calculation of the distance between these atoms $r_{ij}$, and then apply some function $f(r_{ij})$ to that distance. The return value is included in the total sum.  The algorithms presented here can be used for any such distance-dependent function.

For illustrative purposes, the Lennard-Jones potential will be used as an example of the computational power of this new algorithm.  Further extensions and adaptations of the same algorithm to other functions and potentials are discussed in Section \ref{sec:applications}.

The author would like to note there are many common techniques to optimize algorithms, especially nested loops, such as avoiding the repetitive calculation of the same value.  Likewise there are algorithms to avoid round-off error such as the Kahan summation algorithm \cite{Kahan:1965:PRR:363707.363723}.  These common tools are omitted from the algorithms presented here to more clearly show the logic structure, and to more clearly demonstrate what new methods are applied.

\subsection{An Illustrative Example}

%\subsection{LJ Potential}

%%%%%%%%%%%%%%%%%%%%%%%%%%%
%%% put this somewhere
%The distance $d_{j}$ from the origin to any atom $j$ is, of course, $d_{j}=\sqrt{X^2+Y^2+Z^2}$ so Equation \ref{eq:lp} becomes:
%\begin{equation}
%L_p=\sum\limits_{X,Y,Z=-D/2}^{D/2}\frac{1}{(X^2+Y^2+Z^2)^{p/2}}
%\label{eq:Lpwithdist}
%\end{equation}
%%%%%%%%%%%%%%%%%%%%%%%%%%%

The Lennard-Jones potential \cite{Jones636} is a simple but widely-used potential energy formula.  The total potential energy of a crystal with $N$ atoms is described by the sum of Equation (\ref{eq:LJtotalE}) between all pairs of atoms. The constant parameters $\sigma$ and $\epsilon$ are determined from experimental measurements, and $d_{j}$ is the distance from a fixed reference atom to any other atom $j$ as a multiple of the nearest-neighbor distance. 
%It is useful for predicting the structure of crystals such as rare-gas solids which were experimentally seen to take on a different structure than originally expected \cite{RevModPhys.36.748}. A higher precision calculation of the potential energy was required to explain the unexpected result.
\begin{equation}
U_{tot}=2N\epsilon\left[\sum\limits_{j=1}^{\infty}\left(\frac{\sigma}{d_{j}}\right)^{12}-\sum\limits_{j=1}^{\infty}\left(\frac{\sigma}{d_{j}}\right)^{6}\right]
\label{eq:LJtotalE}
\end{equation}

To simplify calculations, it is useful to separate the $d_{j}$ terms and examine them independently:
\begin{equation}
L_p\equiv\sum\limits_{j=1}^{\infty}\left(\frac{1}{d_{j}}\right)^{p}
\label{eq:lp}
\end{equation}

It is seen that Equation (\ref{eq:LJtotalE}) can be determined by first calculating these lattice constants $L_p$ for $p=6$ and $p=12$.  The $p=6$ term
%in Equation (\ref{eq:LJtotalE})
represents the attractive van der Waals force, whereas the Pauli exclusion principle is responsible for the repulsive $p=12$ term. The choice of $p=12$ is not fully motivated from first principles, so it is useful to compute a range of $p$ values.  For $p<4$, the series does not converge \cite{PhysRevB.73.064112}, and for $p>30$, the series is seen to converge to the coordination number of the lattice.  While any real value of $p$ could be computed, this example uses integer values for comparison to other published results which also examine integer values of $p$ \cite{PhysRevB.73.064112,PhysRev.137.A152}.

To achieve a useful value of the lattice constants $L_p$ in Equation (\ref{eq:lp}), the series need only converge to the precision required.  The double-precision float has ${\sim}15$ decimal digits, and is now a very fast variable to use with most modern compilers.  Results have been published for the simple cubic (SC), face-centered cubic (FCC), body-centered cubic (BCC), hexagonal-close-pack (HCP) lattices with up to 15 decimal digits \cite{PhysRevB.73.064112}, but not every term published has actually converged to the precision given, especially for $p<12$.  The diamond (DIA) lattice has been published up to 9 decimal digits \cite{PhysRev.137.A152}, roughly the precision of a 32-bit single-precision float.  To fully demonstrate the power of the algorithms in this work, the Portable, Extensible Toolkit for Scientific Computation (PETSc) \cite{petsc} was used to implement 128-bit floats to push the precision to 32 decimal digits.

\subsection{Brute Force Method}

Consider a SC lattice whose side length is $D$, and whose unit cell has a side length of 1 in arbitrary units. To calculate a distance-dependent function $f(r_{ij})$ over all lattice sites (Equation \ref{eq:lp}), one can set up three nested \textbf{for}--loops to cover a 3-dimensional grid.  Each integer value of the respective loop variables ($X,Y,Z$) represent the coordinates of a particular atom, and sweeping from $-(D/2)$ to $(D/2)$ in all three loops covers all $(D+1)^3$ atoms in the cube.  

The distance $d_{j}$ from the origin to any other atom $j$ is, of course, $\sqrt{X^2+Y^2+Z^2}$ so the program structure then is:
\begin{figure}[h]
\begin{algorithm}[H]
	\caption{Brute Force Method}
	\label{alg:simple}
	\begin{algorithmic}
		\State $L_p$ = 0
		\For{$X\leftarrow -(D/2)$ to $(D/2)$} {
			\For{$Y\leftarrow -(D/2)$ to $(D/2)$} {
				\For{$Z\leftarrow -(D/2)$ to $(D/2)$} {
					\If{$X = 0$ {\bf and} $Y = 0$ {\bf and} $Z = 0$}
					\State Next
					\Else 
					\State $L_p$ += $\frac{1}{(X^2+Y^2+Z^2)^{p/2}}$
					\EndIf
				} \EndFor
			} \EndFor
		} \EndFor
		\State \textbf{return} $L_{p}$
	\end{algorithmic}
\end{algorithm}
\end{figure}

The \textbf{if}--statement is present to avoid the $\frac{1}{0}$ term (at the origin) which would otherwise set $L_{p}$ equal to infinity or $NaN$.  At this point, knowing that there will be $(D+1)^3$ \textbf{if}--statements checked in every run of Algorithm \ref{alg:simple}, it is worth finding how many terms will be necessary for this sum to converge.

\subsection{The Convergent Series} 
Depending on implementation of 128-bit floats \footnote{The IEEE-754 floating point standard defines quadruple-precision, 128-bit floats.  These are implemented differently by various compilers as long-double, \_\_float128, PetscReal, and others.}, these variables yield ${\sim}32$ decimal digits for each term.  Finding where Equation (\ref{eq:lp}) converges then requires additional terms to be equal to or less than $10^{-33}$ (in arbitrary units).  Finding the coordinates of where $L_{p_j}=10^{-33}$ yields little benefit, however, as that is only the value of one such term, and there may be many such terms at that distance.

\begin{figure}[ht]
	\caption{Average value of terms added to $L_6$ from the face at some fixed $R$ versus the distance and number of terms added to $L_6$ at that face.}
	\includegraphics[width=\linewidth]{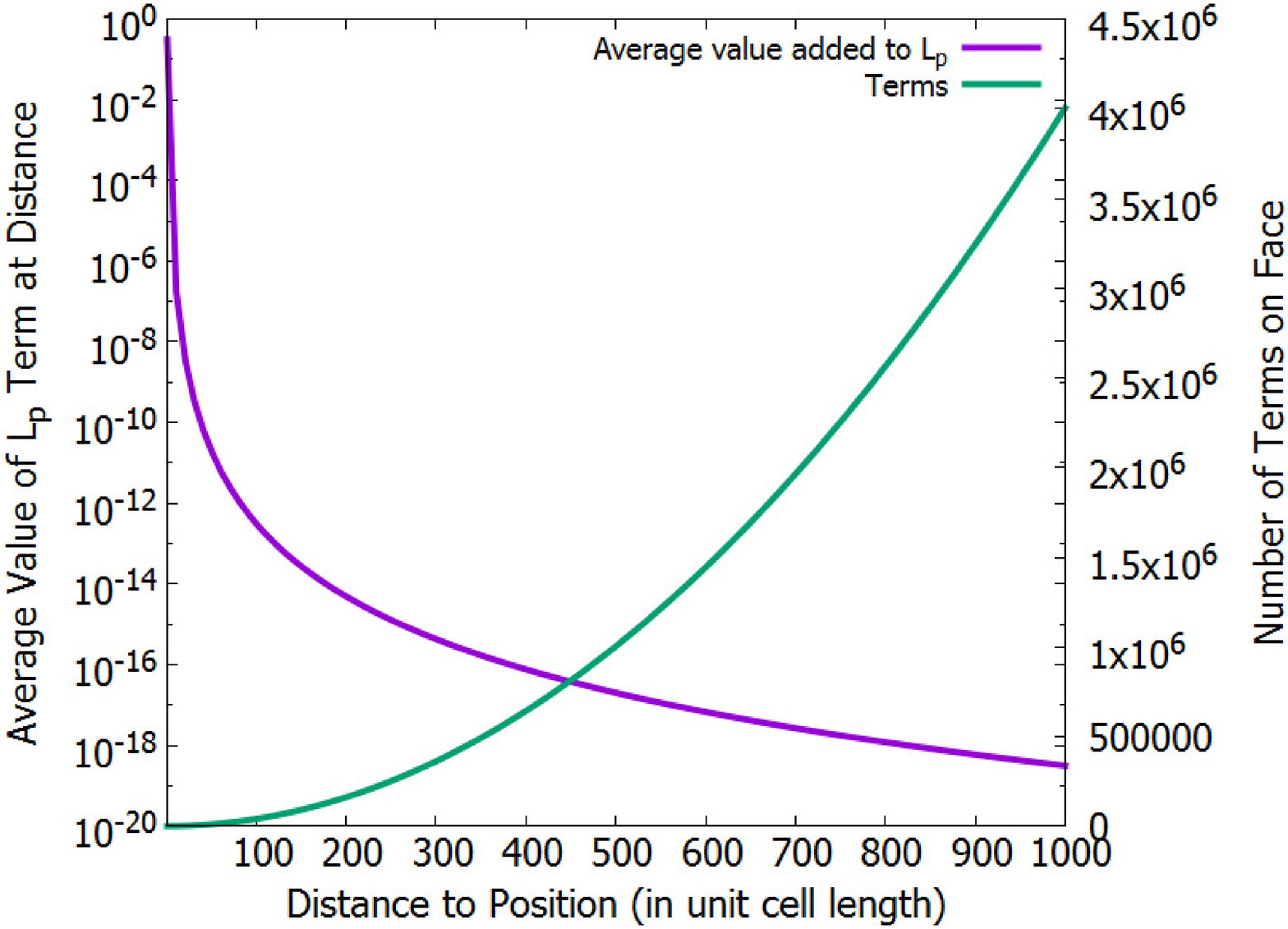}
	\label{im:termsFacePlot}
\end{figure}

For example, say $L_{p_j}=\frac{1}{(X_{j}^2+Y_{j}^2+Z_{j}^2)^{p/2}}=10^{-33}$ for atom $j$ at $(X_j,Y_j,Z_j)$, and say $Y_j=Z_j=0$ for simplicity.  In the brute force method described above, the algorithm will still be computing approximately $R^2$ more terms for the face at $X_j=R$.  Moreover, there will be six such faces to add to the total sum.  Higher distances decrease the value of each $L_{p_j}$ term, but there are more terms to the total sum at some fixed $R$, slowing down the convergence of the series with increasing distance (Figure \ref{im:termsFacePlot}).  One can calculate the total amount added to $L_p$ from adding one layer at a fixed $R$ distance, showing the slowness of convergence. For $L_6$, the total value added from one layer at distance $R$ goes as $1/R^{4}$ (Equation \ref{eq:FaceValues}).  This is determined by integrating Equation (\ref{eq:lp}) with respect to $Y$ and $Z$ for $p=6$ and $X=R$. That result is multiplied by 6 for symmetry.  While an exact result requires the actual summation in Equation (\ref{eq:lp}), this result is useful for determining how many terms are requires for convergence to a particular precision.
\\
\begin{equation*}
Sum_{p,face@R}\propto\frac{1}{R^{(p-2)}}
\label{eq:GenFaceValues}
\end{equation*}
\begin{equation}
Sum_{6,face@R}=6\times\frac{2+15\sqrt{2}\textrm{ArcCot}\sqrt{2}}{12 R^4}\approx\frac{7.52815}{R^4}
\label{eq:FaceValues}
\end{equation}

\begin{figure}[ht]
	\caption{Average value of terms added to $L_p$ across the face at some fixed $R$ versus the distance $R$. One can draw a horizontal line across the graph at the desired precision on the vertical axis.  Where that line intersects each $p$ function will be approximately the distance required to converge the sum at that precision.}
	\includegraphics[width=\linewidth]{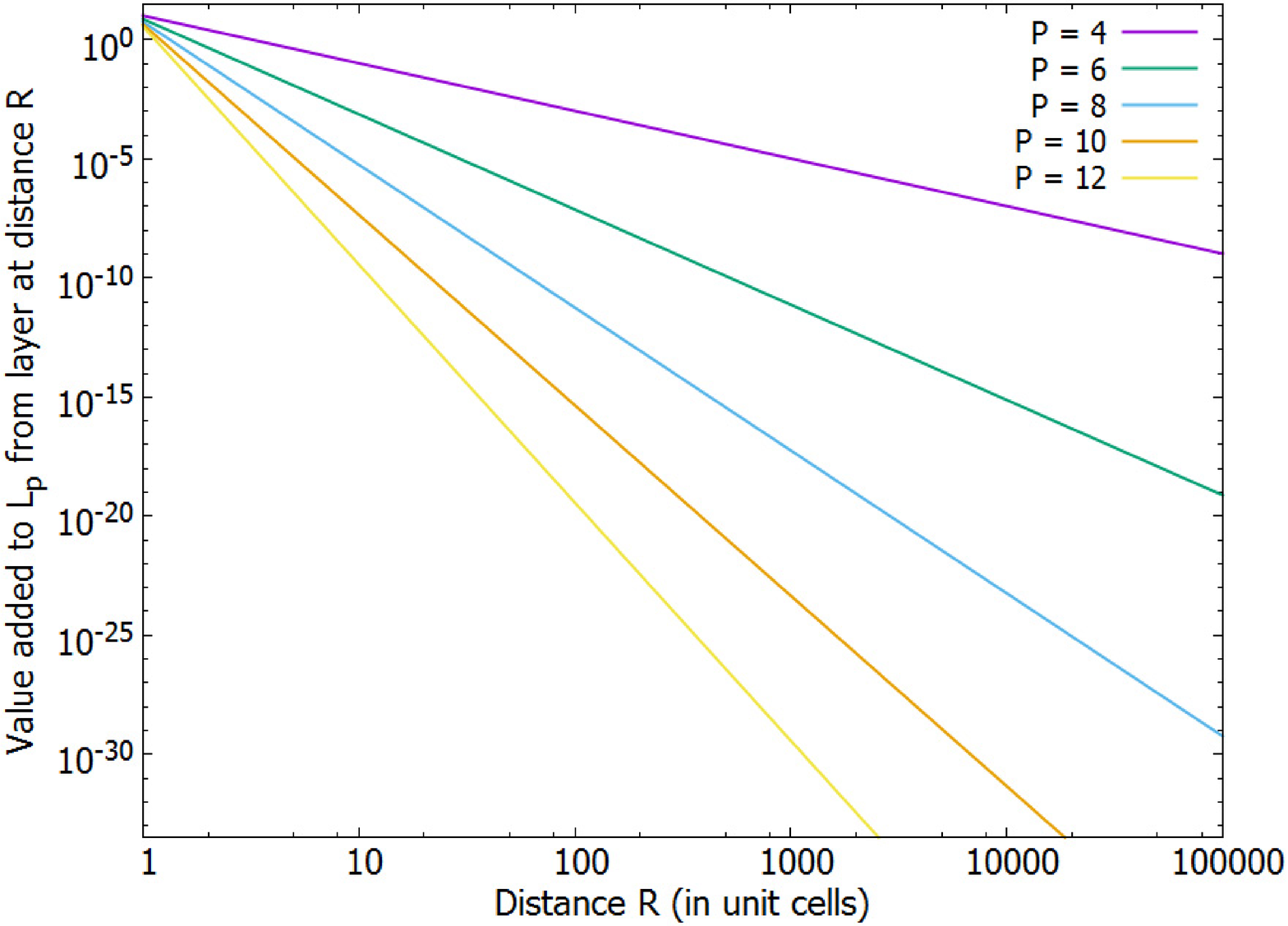}
	\label{im:conv}
\end{figure}

The convergence of Equation (\ref{eq:lp}) is much faster for higher values of $p$ (Figure \ref{im:conv}) but presents a significant computational challenge for low $p$. Converging to any desired precision at low $p$ will then require finding fast algorithms that will capitalize on efficiency, parallelism, and any inherent symmetries in the crystal lattice.

\subsection{Finding Speedup}
\subsubsection{Avoiding Unnecessary Operations}
\begin{figure}[ht]
	\caption{The six volumes to loop over, automatically avoiding the unit cell at the origin.  The red regions indicate the 2-dimensional face planes, and the green regions are the 1-dimensional axes.}
	\includegraphics[width=\linewidth]{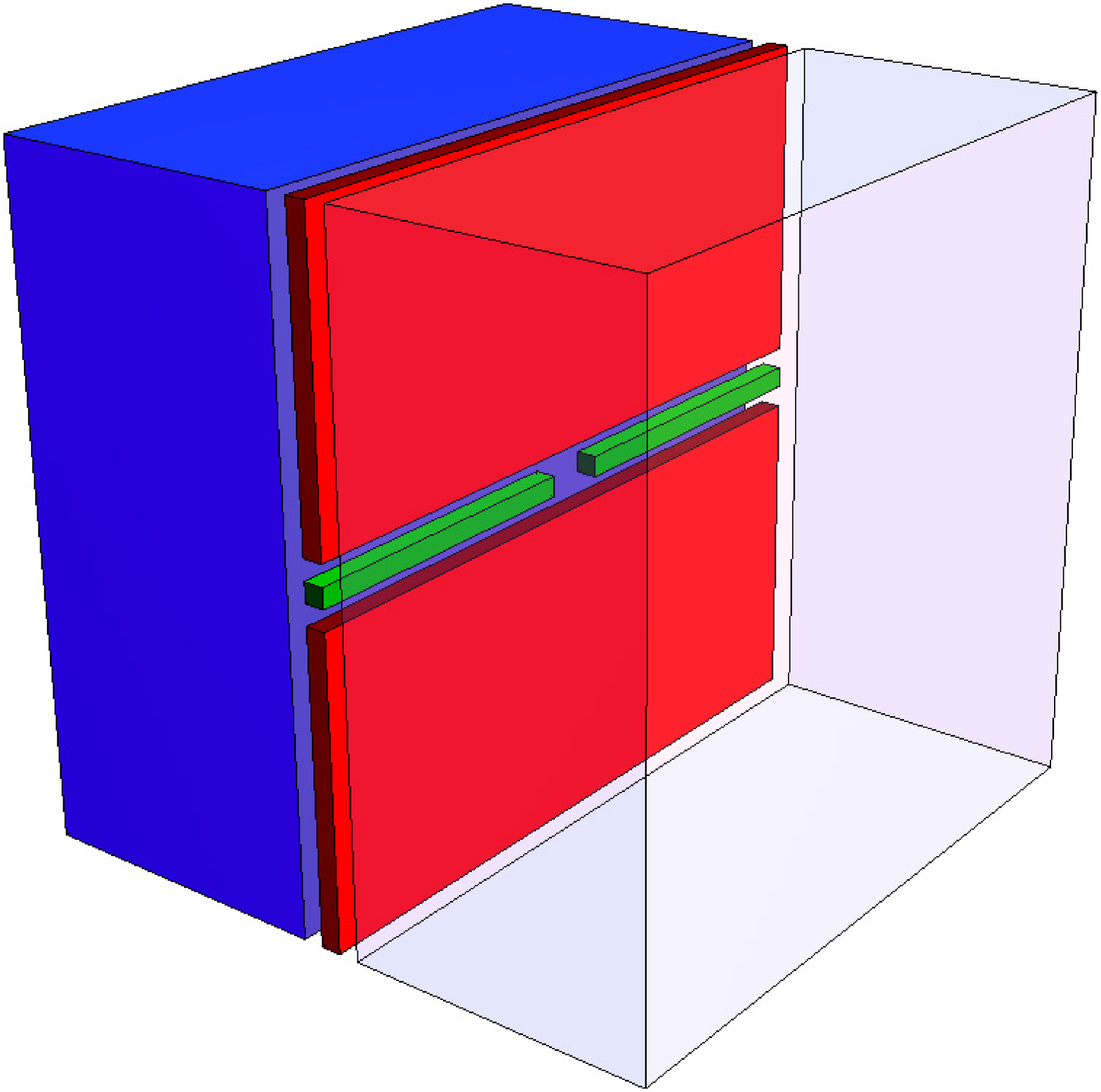}
	\label{im:split1}
\end{figure}

In the simple case of Algorithm \ref{alg:simple}, the $(D+1)^3$ \textbf{if} statements can be avoided by structuring the program to calculate different regions of the same cube, none of which contain the (0,0,0) position (Figure \ref{im:split1}).  There are now six regions to consider: two rectangular parallelepipeds, two planes, and two lines.  The loops for these regions are executed in serial (Algorithm \ref{alg:BrokenDown}).

\begin{figure}
\begin{algorithm}[H]
	\caption{Broken Down into Six Separate Regions}
	\label{alg:BrokenDown}
	\begin{algorithmic}
		\State $L_p$ = 0
		\State // Cube Volumes
		\For{$X\leftarrow -(D/2)$ to $(D/2)$}{
			\For{$Y\leftarrow -(D/2)$ to $(D/2)$}{
				\For{$Z\leftarrow -(D/2)$ to $-1$}
				\State $L_p$ += $\frac{1}{(X^2+Y^2+Z^2)^{p/2}}$
				\EndFor
			} \EndFor
		} \EndFor 
		\For{$X\leftarrow -(D/2)$ to $(D/2)$}{
			\For{$Y\leftarrow -(D/2)$ to $(D/2)$}{
				\For{$Z\leftarrow 1$ to $(D/2)$}
				\State $L_p$ += $\frac{1}{(X^2+Y^2+Z^2)^{p/2}}$
				\EndFor
			} \EndFor
		} \EndFor
		\State // Faces @ $Z = 0$
		\For{$X\leftarrow -(D/2)$ to $(D/2)$}{
			\For{$Y\leftarrow -(D/2)$ to $-1$}
			\State $L_p$ += $\frac{1}{(X^2+Y^2)^{p/2}}$
			\EndFor
		} \EndFor
		\For{$X\leftarrow -(D/2)$ to $(D/2)$}{
			\For{$Y\leftarrow 1$ to $(D/2)$}
			\State $L_p$ += $\frac{1}{(X^2+Y^2)^{p/2}}$
			\EndFor
		} \EndFor
		\State // Axes @ $Y = 0$ and $Z = 0$
		\For{$X\leftarrow -(D/2)$ to $-1$}
		\State $L_p$ += $\frac{1}{X^{p}}$
		\EndFor
		\For{$X\leftarrow 1$ to $(D/2)$}
		\State $L_p$ += $\frac{1}{X^{p}}$
		\EndFor
		\State return $L_{p}$
	\end{algorithmic}
	
\end{algorithm}
\end{figure}

\subsubsection{Parallelization}
Since each individual $L_{p_i}$ value is independent of every other $L_{p_j}$, Algorithm \ref{alg:BrokenDown} is an excellent candidate for parallelization via MPI \cite{MPIcite}. The parallelization of these nested \textbf{for}--loops, however, requires the following careful prescription such that each thread does approximately the same amount of work, and the entire 3-dimensional grid of lattice points is covered.  For $NumProcs$ threads, one cannot simply set thread number $MyID$ to cover a range of $(D/NumProcs)$ in $(X/Y/Z)$ as can be trivially done in the case of a 1-dimensional array.  Instead, the original cube from Algorithm \ref{alg:simple} is broken down into $NumProcs$ inter-penetrating cubes with a different basis.  This allows every thread to compute $\frac{(D+1)^3}{NumProcs}$ elements of $L_p$, the results of which can be combined at the end of the algorithm. The integer basis of each new lattice is computed as follows:
\begin{equation}
Basis=Floor(\sqrt[3]{NumProcs})
\label{eq:BasisIC}
\end{equation}
The initial $(X,Y,Z)$ position of each thread is:
\begin{align}\label{eq:xiyizi}
X_i&=(MyID\%Basis)+(D/2)\nonumber\\
Y_i&=(\textrm{Floor}\left[\frac{MyID}{Basis}\right]\%Basis)+(D/2) \\
Z_i&=(\textrm{Floor}\left[\frac{MyID}{Basis^2}\right]\%Basis)+(D/2) \nonumber
\end{align}

Fortunately, only one thread (hereafter the \it{origin thread}\rm) will pass through the $(0,0,0)$ position.  All other threads can execute a fast triple-nested \textbf{for}--loop (Algorithm \ref{alg:Parallel}, Figure \ref{im:OneThread}), and the origin thread will execute a slightly modified version of Algorithm \ref{alg:BrokenDown}. The origin thread is identified as:
\begin{align*}
k=(D/2)\%Basis\end{align*}
\begin{align*}
OriginThreadNum=k*Basis^2+k*Basis+k 
\end{align*}

A simple MPI summation is performed at the end of the program, and the result is returned.
\begin{figure}
\begin{algorithm}[H]
	%\TitleOfAlgo{3}
	\caption{Invoked in Parallel \\(Threads other than origin thread)}
	\label{alg:Parallel}
	\begin{algorithmic}
		\State $L_p$ = 0
		\State ... set $X_i$, $Y_i$, $Z_i$...
		%\tcp{Cube Volume 1}
		\For{$X\leftarrow X_i$ to $-(D/2)$ in steps of $-Basis$}{
			\For{$Y\leftarrow Y_i$ to $-(D/2)$ in steps of $-Basis$}{
				\For{$Z\leftarrow Z_i$ to $-(D/2)$ in steps of $-Basis$}
				\State $L_p$ += $\frac{1}{(X^2+Y^2+Z^2)^{p/2}}$\;	
				\EndFor
			}\EndFor
		}\EndFor
		\State ... MPI summation ...
		\State return $L_{p,total}$
	\end{algorithmic}
	
\end{algorithm}
\end{figure}

\begin{figure}[ht]
	\caption{Example of atom sites in the SC lattice looped over by a single thread (blue) for an 8-thread invocation of Algorithm \ref{alg:Parallel}. The remaining green sites are divided among the other seven threads.  In a real lattice, the spheres should be uniform and expanded to fill the maximum volume possible, but are shown with different sizes here for clarity.}
	\includegraphics[width=\linewidth]{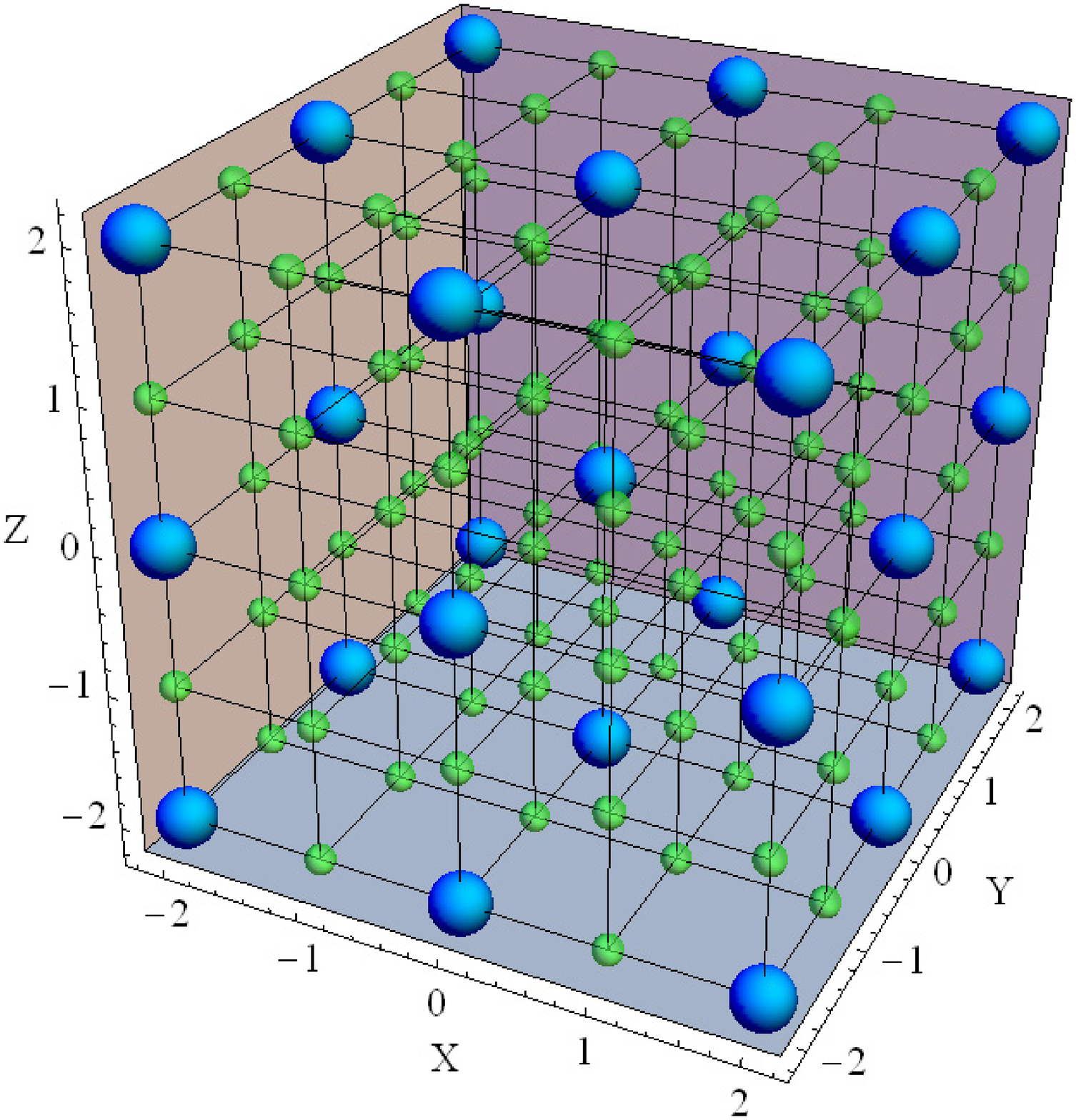}
	\label{im:OneThread}
\end{figure}

%The speedup in the evolution from Algorithm \ref{alg:simple} to \ref{alg:BrokenDown} to \ref{alg:Parallel} is XXX (Figure XXX showing speedup).
One caveat with this prescription is that it requires $NumProcs$ to have an integer cube root.  On small clusters with a limited number of threads, this can prevent the full utilization of this method, but even consumer processors are widely available in 8-core (or more) configurations which is the minimum required. More flexible methods not requiring a cubic number of threads are possible, but come at a performance cost.  The cluster used for this example (Southern Methodist University's \mbox{ManeFrame}) has over 1,100 CPU nodes available, each with eight cores, making over 8,800 simultaneous threads possible, erasing the need for programming more flexible methods. 

\subsubsection{Exploiting Symmetry}
\begin{figure}[ht]
	\caption{SC unit cells are shown along the plane at $X=0$.  The atoms (shown in yellow; not to scale) can be shared between different octants at such a plane.}
	\includegraphics[width=\linewidth]{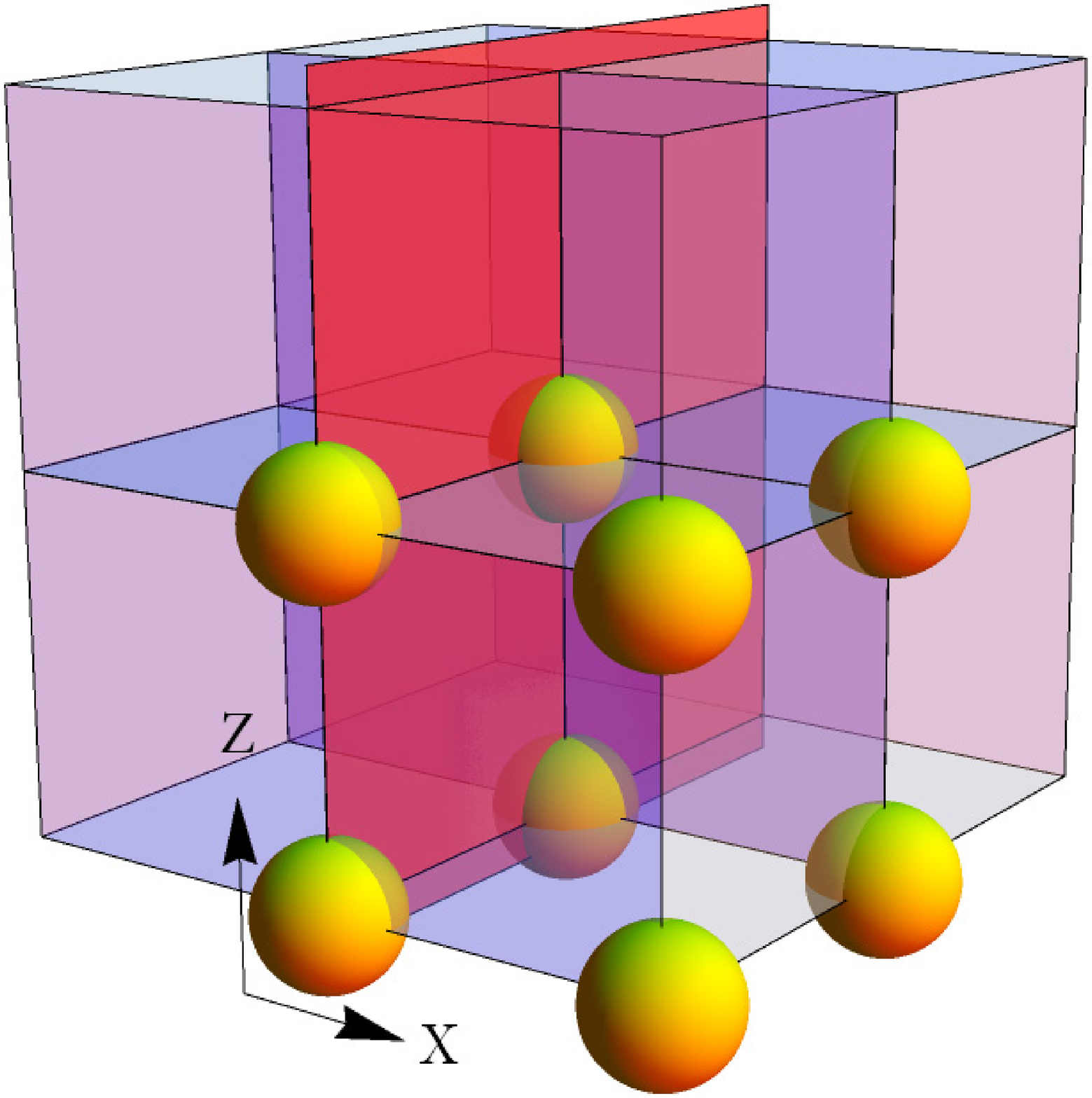}
	\label{im:cubes}
\end{figure}

\begin{figure}[ht]
	\caption{The shared volumes of the planes and axes between the octants}
	\includegraphics[width=\linewidth]{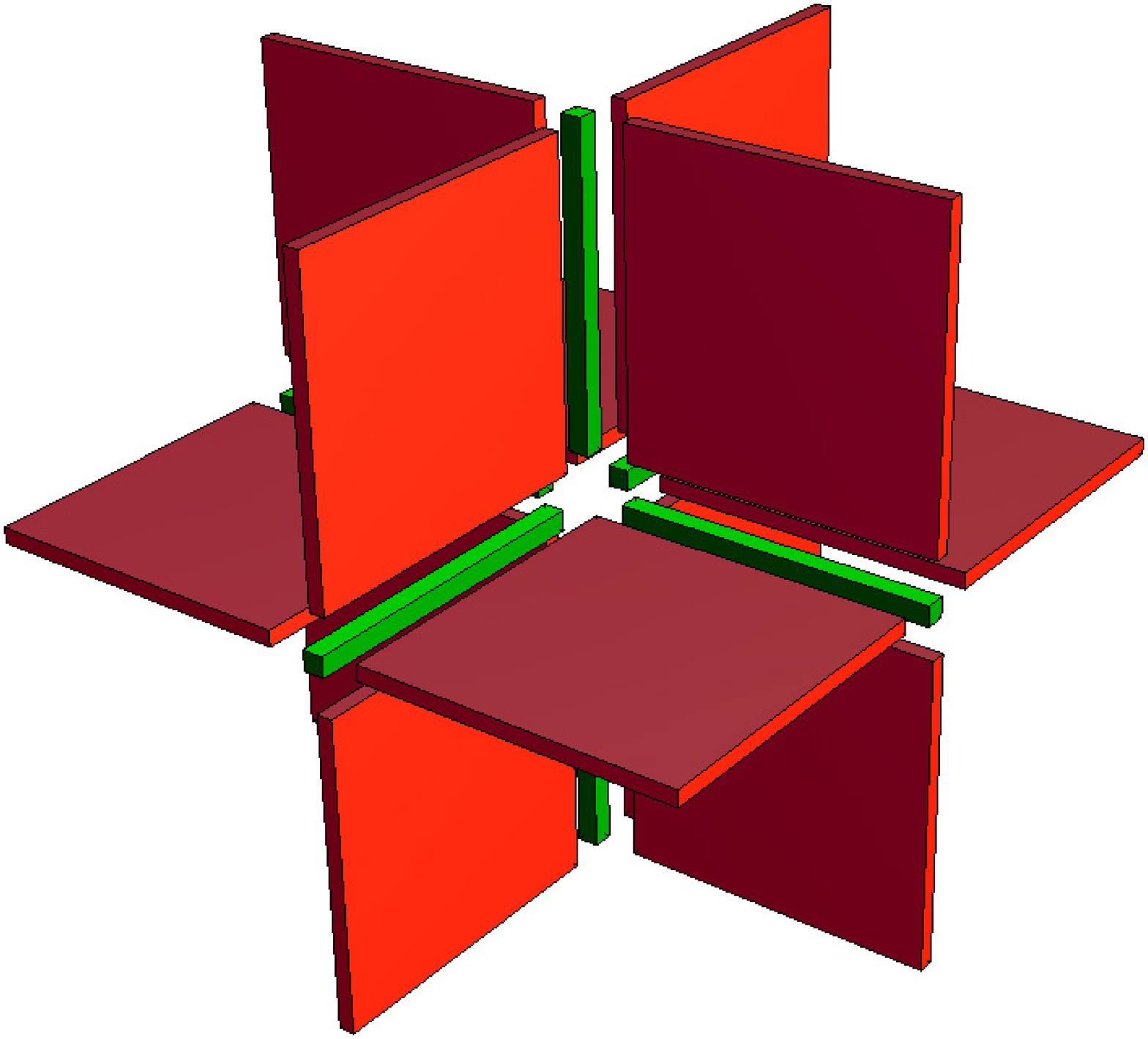}
	\label{im:planesAxes}
\end{figure}

In the case of the SC lattice, the calculation of $L_p$ can be shortened by considering that the cube is made of eight identical, smaller pieces corresponding to each octant.  Therefore a speedup of almost eightfold can be found by calculating only one of these octants and multiplying the end result.  However, the algorithmic range of each octant is not as obvious as it seems.  There are unit cells along the planes between octants whose atoms need to have their contributions handled carefully as some of the atoms sit astride different octants (Figure \ref{im:cubes}), and likewise for cells along the axes. For unit cells immediately adjacent other octants, consider these as being in separate volumes called the axis or face volumes (for cells touching the axes or faces between octants, respectively) as in Figure \ref{im:planesAxes}). The remaining cells are considered to be in one of eight cubic volumes spanning the rest of each octant. Therefore, in the entire lattice, there are eight cubic volumes, 12 face volumes, and six half-axis volumes.  In the case of the SC lattice, one need only calculate the sum of a single cubic volume $L_{cube}$, a single face $L_{face}$, and a single half-axis $L_{axis}$ to determine $L_p$ (Equation \ref{eq:SymCalc}).  

The values $1.5$ and $0.75$ arise in Equation (\ref{eq:SymCalc}) from the fact that there are 12 faces and six half-axes that should contribute equally to each of the eight octants, so $\frac{12}{8}$ for the faces, and $\frac{6}{8}$ for the axes.  The new serial algorithm, which automatically avoids the origin, can be written compactly as in Algorithm \ref{alg:Sym}.
\begin{equation}
L_p=\left[L_{cube}+(1.5*L_{face})+(0.75*L_{axis})\right]*8
\label{eq:SymCalc}
\end{equation}
\begin{figure}
\begin{algorithm}[H]
	%\TitleOfAlgo{4}
	\caption{Symmatrized}
	\label{alg:Sym}
	\begin{algorithmic}
		\State $L_{cube} = L_{face} = L_{axis}$ = 0
		%... set $X_i$, $Y_i$, $Z_i$ in farthest-first method...\\
		%\tcp{Cube Volume}
		\For{$X\leftarrow (D/2)$ to $1$}
		\For{$Y\leftarrow (D/2)$ to $1$}
		\For{$Z\leftarrow (D/2)$ to $1$}
		\State $L_{cube}$ += $\frac{1}{(X^2+Y^2+Z^2)^{p/2}}$
		\EndFor
		\State $L_{face}$ += $\frac{1}{(X^2+Y^2)^{p/2}}$
		\EndFor
		\State $L_{axis}$ += $\frac{1}{X^p}$
		\EndFor
		\State return $\left[L_{cube} + (1.5\times L_{face}) + (0.75\times L_{axis})\right]*8$
	\end{algorithmic}
	
\end{algorithm}
\end{figure}

To parallelize this, the basis is calculated as before, but the same basis does not hold for calculation of the face or axis (2- and 1-dimensional arrays, respectively). The calculation of those bases is shown in Algorithm \ref{alg:FaceAxisBasis}.
%and the ranges can be recalculated in the farthest-first method as in Equation \ref{eq:xiyizi}, substituting $Basis$ for $Basis_f$ and $Basis_a$, respectively.
%\begin{align}
%X_i&=(MyID\%Basis)+(D/2) \nonumber\\
%Y_i&=(\textrm{Floor}\left[\frac{MyID}{Basis}\right]\%Basis)+(D/2) \\
%Z_i&=(\textrm{Floor}\left[\frac{MyID}{Basis^2}\right]\%Basis)+(D/2)\nonumber
%\end{align}
When looping through each volume (cube, face, or axis as shown in Algorithm \ref{alg:ParallelSym}), the values of ($X_i,Y_i,Z_i$) must be calculated relative to the appropriate basis for that volume (Equation \ref{eq:xiyizi} with $Basis$, $Basis_f$ or $Basis_a$, as needed). Note that the step sizes must be negative since the initial positions are set at points away from the origin. The new algorithm is computed by all threads as there is no need to find an origin thread.

\begin{figure}
\begin{algorithm}[H]
	\caption{Calculation of Basis}
	\label{alg:FaceAxisBasis}
	\begin{algorithmic}
		%\TitleOfAlgo{2}
		\State $T=NumProcs$
		\While{($\sqrt{T}-$Floor($\sqrt{T}$))\textgreater$0$}
		\State $T--$
		\EndWhile
		\State $Basis_{f}=$Floor($\sqrt{T}$)
		\State $Basis_{a}=NumProcs-T$
	\end{algorithmic}
\end{algorithm}
\end{figure}

\begin{figure}
\begin{algorithm}[H]
	%\TitleOfAlgo{5}
	\caption{Parallelized \& Symmatrized}
	\label{alg:ParallelSym}
	\begin{algorithmic}
	\State $L_{cube} = L_{face} = L_{axis}$ = 0
	%... set $X_i$, $Y_i$, $Z_i$, and bases...\\
	\State // Cube Volume
	\State ... calculate $X_i$, $Y_i$, $Z_i$ relative to $Basis$ ...
	\For{$X = X_i$ to $X>0$ in steps of $-Basis$}
		\For{$Y = Y_i$ to $Y>0$ in steps of $-Basis$}
			\For{$Z = Z_i$ to $Z>0$ in steps of $-Basis$}
				\State $L_{cube}$ += $\frac{1}{(X^2+Y^2+Z^2)^{p/2}}$
			\EndFor
		\EndFor
	\EndFor
	\State // Face \& Axis
	\If{$Basis_a > 0$}
		\If{$MyID < T$}
			\State ... calculate $X_i$, $Y_i$ relative to $Basis_{f}$ ...
			\For{$X = X_i$ to $X>0$ in steps of $-Basis_{f}$}
				\For{$Y = Y_i$ to $Y>0$ in steps of $-Basis_{f}$}
					\State $L_{face}$ += $\frac{1}{(X^2+Y^2)^{p/2}}$
				\EndFor
			\EndFor
		\Else
			\State ... calculate $X_i$ relative to $Basis_{a}$ ...
			\For{$X = X_i$ to $X>0$ in steps of $-Basis_{a}$}
				\State $L_{axis}$ += $\frac{1}{X^p}$
			\EndFor
		\EndIf

	\Else
		\State ... calculate $X_i$, $Y_i$ relative to $Basis_{f}$ ...
		\For{$X = X_i$ to $X>0$ in steps of $-Basis_{f}$}
			\For{$Y = Y_i$ to $Y>0$ in steps of $-Basis_{f}$}
				\State $L_{face}$ += $\frac{1}{(X^2+Y^2)^{p/2}}$
			\EndFor
		\EndFor
		\State ... set $X_i$ as $ProcID + D$ ...
		\For{$X = X_i$ to $X>0$ in steps of $-NumProcs$}
			\State $L_{axis}$ += $\frac{1}{X^p}$
		\EndFor
	\EndIf
	\State ... MPI summation ...
	\State return $\left[L_{cube} + (1.5\times L_{face}) + (0.75\times L_{axis})\right]*8$
	\end{algorithmic}
\end{algorithm}
\end{figure}
\subsubsection{Extending the Exploitation of Symmetry: BCC and FCC}
For the BCC and FCC lattices, the same exploitation of octants can be used, but with special handling:  The lattices must be thought of as an SC lattice with two and four basis atoms, respectively.  The \textbf{for}--loop variables now indicate the coordinates of the new conventional unit cells, instead of just the atoms.  The nearest-neighbor distance must be normalized properly to this new conventional unit cell (Table \ref{table:NormFactors}), and the first triple-nested \textbf{for}--loop in Algorithm \ref{alg:FCC} can be computed similarly to the SC case, with the additional basis atom(s) added at each unit cell location.  However, the face- and axis-cells are handled uniquely.
\begin{figure}
\begin{algorithm}[H]
	\caption{FCC - Symmatrized}
	\label{alg:FCC}
	\begin{algorithmic}
	\State $L_{cube} = L_{face} = L_{axis}$ = 0
	\State $n = 2.0$ // n = Normalization factor
	\State // Basis atom offsets
	\State $b2x = 0.5; b2y = 0.5; b2z = 0.0$
	\State $b3x = 0.0; b3x = 0.5; b3z = 0.5$
	\State $b4x = 0.5; b4y = 0.0; b4z = 0.5$
	%... set $X_i$, $Y_i$, $Z_i$ in farthest-first method...\\
	%\tcp{Cube Volume}
	\For{$X\leftarrow (D/2)$ to $1$}
		\For{$Y\leftarrow (D/2)$ to $1$}
			\For{$Z\leftarrow (D/2)$ to $1$}
				\State // First basis atom
				\State $R = (X^2+Y^2+Z^2)*n$
				\State $L_{cube}$ += $\frac{1}{R^{p/2}}$
				\State // Second basis atom
				\State $R = \left((X+b2x)^2+(Y+b2y)^2+(Z+b2z)^2\right)*n$\
				\State $L_{cube}$ += $\frac{1}{R^{p/2}}$
				\State ... similarly for the other basis atoms ...
			\EndFor
			\State $R = (X^2+Y^2)*n$
			\State $L_{face}$ += $\frac{1}{R^{p/2}}$
			\State ... then count basis atom 2 once ...
			\State ... then count basis atoms 3 \& 4 twice ...
		\EndFor
		\State $R = (X^2)*n$
		\State $L_{axis}$ += $\frac{1}{R^{p/2}}$
		\State ... then count basis atom 2 twice ...
		\State ... then count basis atom 3 four times ...
		\State ... then count basis atom 4 once ...
	\EndFor
	\State return $\left[L_{cube} + (1.5\times L_{face}) + (0.75\times L_{axis})\right]*8$
	\end{algorithmic}
\end{algorithm}
\end{figure}

The multiple counts of basis atoms in Algorithm \ref{alg:FCC} are due to the way they are ``shared" between the octants of the broken up cube.  Along the face or axis, there are cells where an atom sits astride the plane separating the octants.  These atoms should only be counted once.  However, there are other basis atoms in the volume of the unit cells immediate next to these planes (or around the axes) which need to be effectively counted once, but since the number of faces and axes that are shared is different from the number of octants, they need to be counted with special weights.

\begin{figure}[]
	\caption{Conventional FCC unit cells along the X-Y plane (red).  For face terms, the shared atoms (yellow spheres) are counted once in Algorithm \ref{alg:FCC}, whereas the mirrored atoms (solid blue spheres) must be counted twice to make full use of symmetry. The faded blue spheres indicate which atoms are mirroring those indicated by the solid blue atoms. Gray spheres indicate other FCC atoms in other unit cells along the red plane.}
	\includegraphics[width=\linewidth]{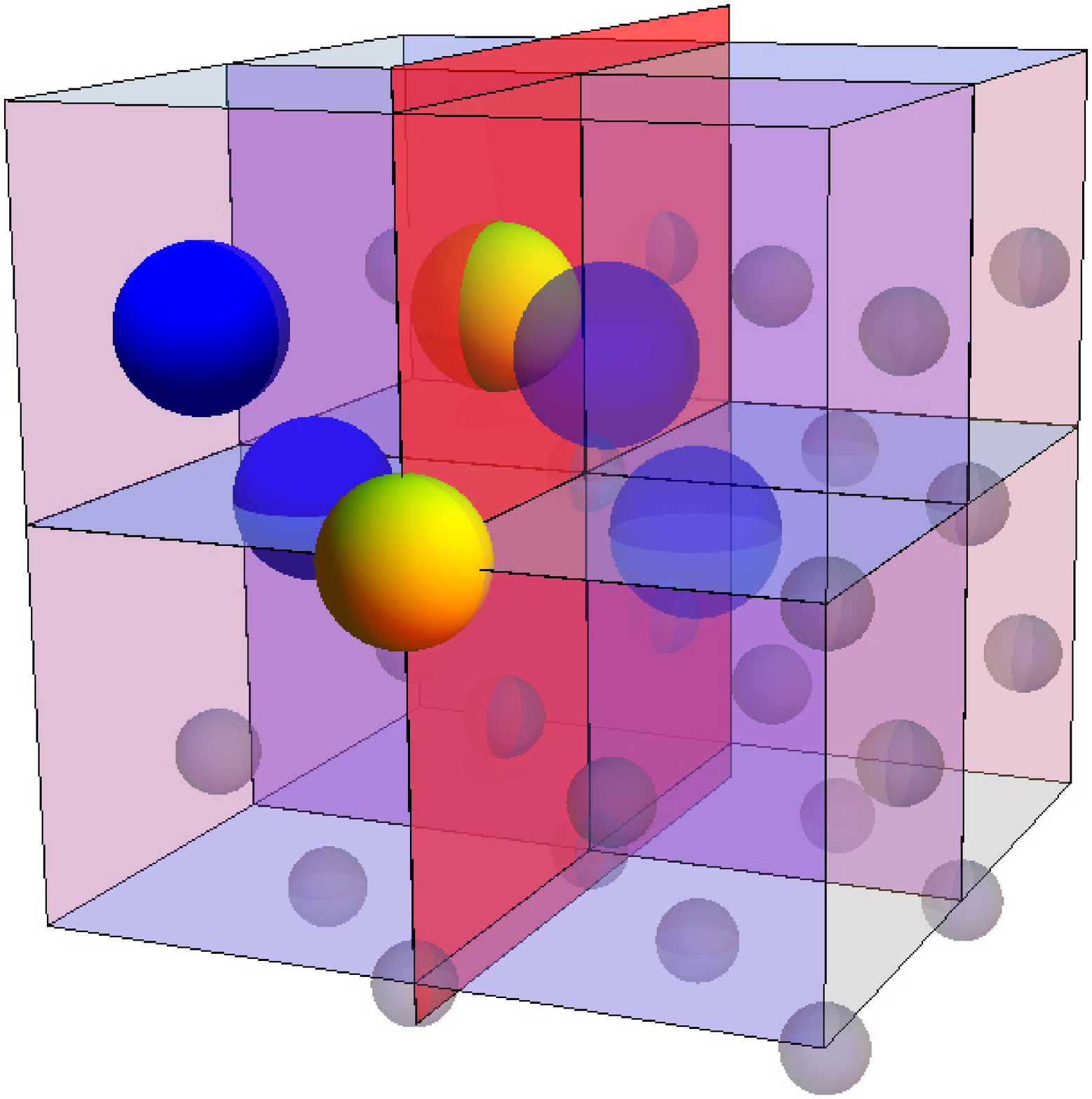}
	\label{im:Mirrored}
\end{figure}

Consider the unit cells spanning the X-Y plane of a single octant.  To minimize calculations, it is possible to compute only terms from these cells and use symmetry to apply the results to the Y-Z and X-Z planes.  In the case of the SC lattice, all atoms sit astride the axes and faces evenly, so no special counting or weighing is needed.  In the FCC case, two of the atoms sit evenly across the X-Y plane, and two are mirrored (Figure \ref{im:Mirrored}).  These mirrored atoms must be counted twice as in Algorithm \ref{alg:FCC}, and then the total contribution from the face can be added to the sum.

Similar to the faces, the algorithm need only compute the unit cells along a single half-axis.  In the case of cells along the X-axis, the first basis atom is counted once as it is shared evenly between all four unit cells located around around the axis.  The second and fourth basis atoms (locations described in Algorithm \ref{alg:FCC}) are shared evenly between two cells, and mirrored across one plane, so they are each counted twice.  The third basis atom is mirrored in all four unit cells along the axis, so it is counted four times.  

In the BCC structure, the same mirroring principle applies: the second basis atom is counted once for cells in the 3-dimensional volume spanned, twice for cells along the face, and four times for cells along the axis. 

\begin{table}[h]
	\begin{tabular}{ccc}\hline
		\Tstrut Lattice & Basis & Normalization\\
		& Atoms & Factor \\\hline
		\Tstrut SC & $1$ & $1$\\[1.9pt]
		BCC  & $2$ & $\frac{4}{3}$\\[1.9pt]
		FCC & $4$ & $2$\\[1.9pt]
		DIA & $12$ & $\frac{16}{3}$\Bstrut\\\hline
	\end{tabular}
	\caption{Number of basis atoms and normalization factors in the conventional unit cells for each lattice.\label{table:NormFactors}}
\end{table}

\subsubsection{Extending the Exploitation of Symmetry: DIA}
\begin{figure}[ht]
	\caption{Two conventional cells of the DIA lattice are shown. Distances to the tetrahedral atoms are not the same when rotating 90 degrees about any point. The blue spheres are the FCC-like basis atoms, and the red spheres indicate the tetrahedral atoms. The sizes of the spheres are not indicative of the sizes of the atoms at these sites.}
	\includegraphics[width=\linewidth]{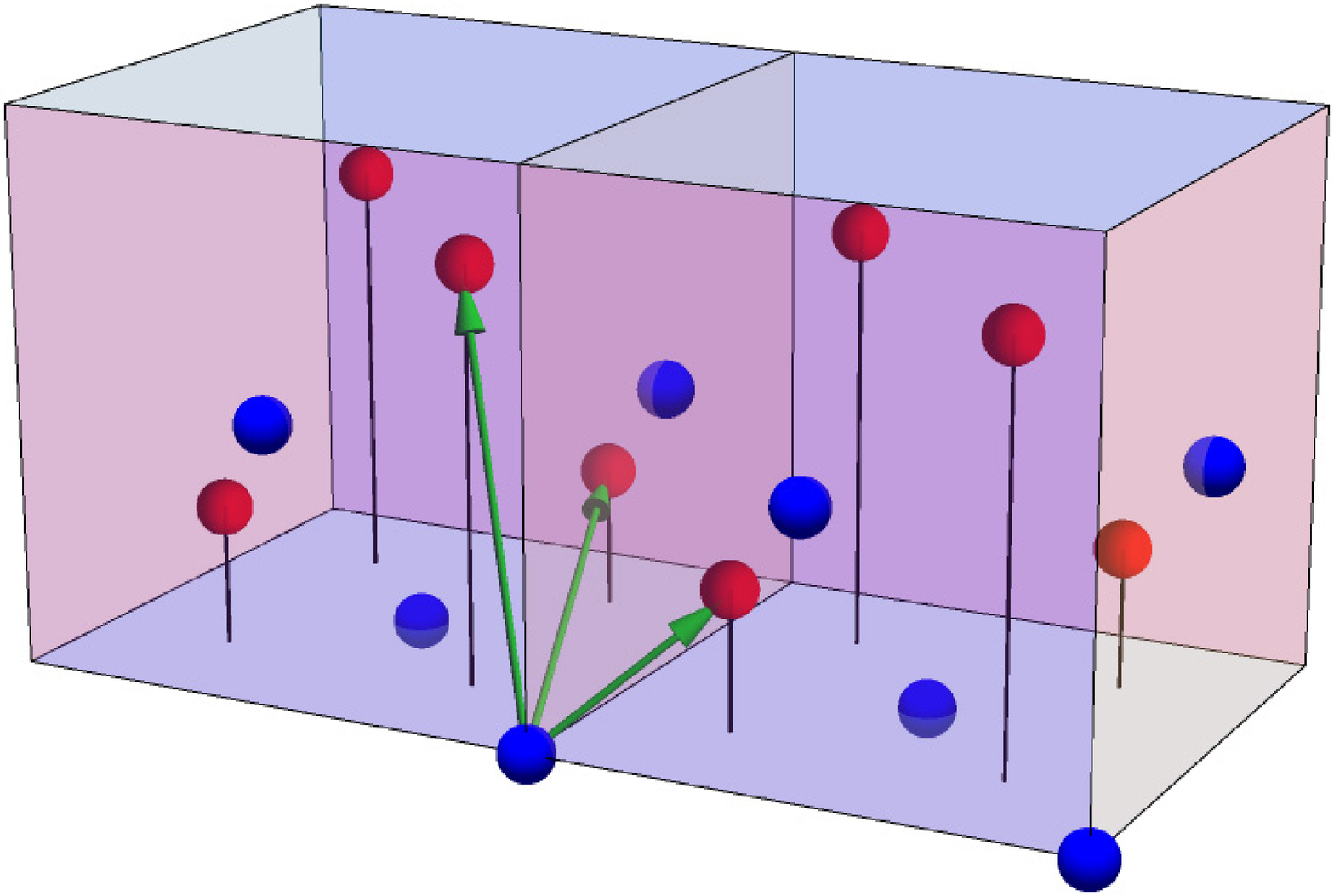}
	\label{im:DIAdist}
\end{figure}

For diamond, the conventional unit cell is essentially an FCC conventional cell with the addition of four more basis atoms within the volume of the cell at the tetrahedral positions.  The algorithm requires further special handling due to the asymmetry of the tetrahedral positions across one axis.  In the BCC and FCC cases, the symmetry between octants obeyed rotational symmetry in that rotating the view 90 degrees about any axis resulted in viewing the exact same configuration of atoms.  However, the diamond lattice does not have this symmetry.  When rotating 90 degrees, the tetrahedral atoms now appear at different distances (Figure \ref{im:DIAdist}).

\begin{figure}[ht]
	\caption{The 12-basis-atom conventional unit cell for calculations involving the diamond lattice. The orange spheres indicate the additional tetrahedral atoms.}
	\includegraphics[width=\linewidth]{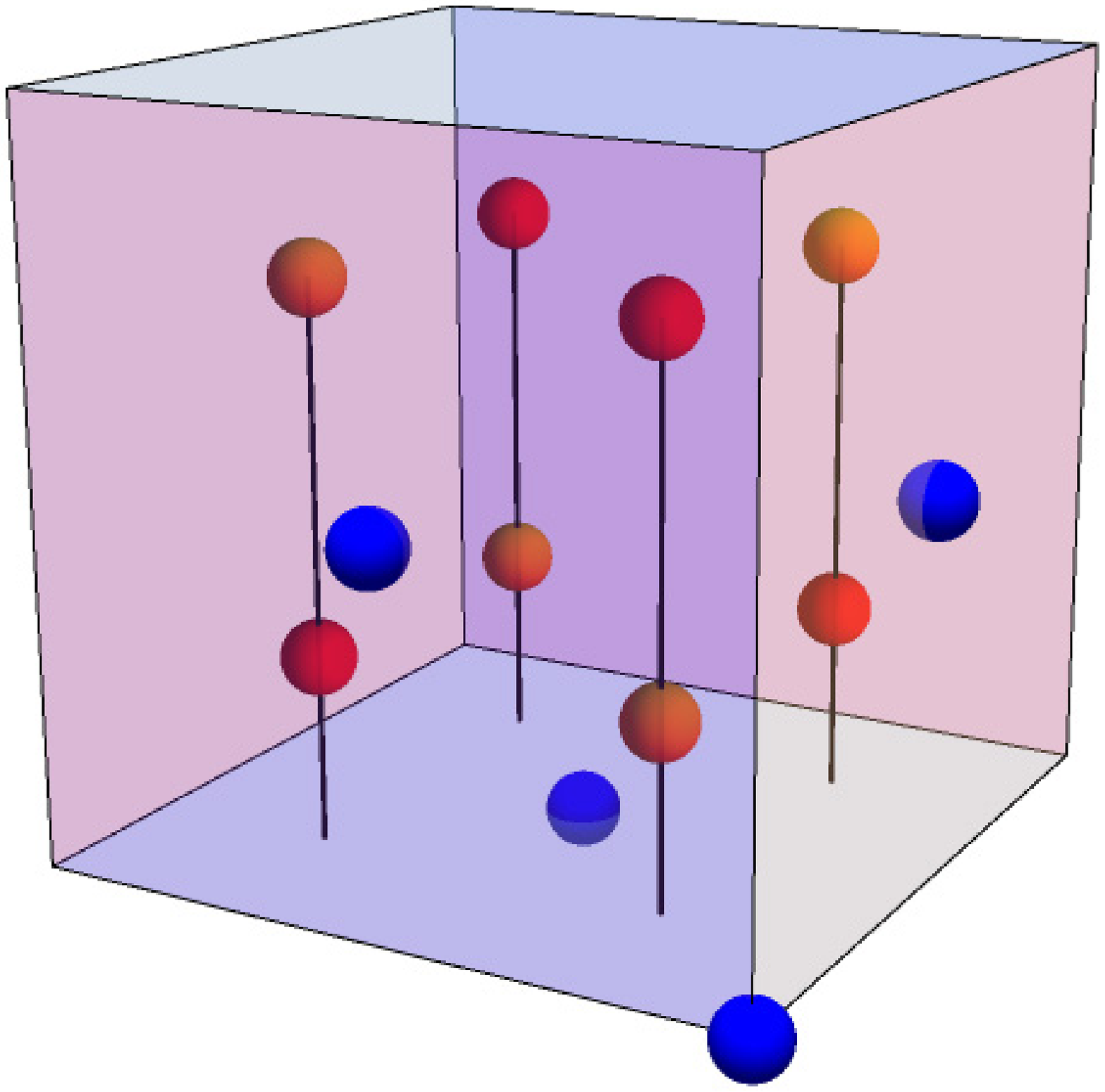}
	\label{im:DIA12}
\end{figure}

Rather than settling for only four-fold speedup in symmetrization, one can fashion a new conventional unit cell that, while physically unrealistic,  presents the same mathematical results as a real DIA lattice for this calculation.  The new conventional unit cell has 12 basis atoms where four are the usual FCC-like atoms, four are the original tetrahedral atoms, and an additional four atoms occupy the location of where the tetrahedral atoms would appear to be if the viewer rotates 90 degrees (Figure \ref{im:DIA12}).

All of the tetrahedral atom contributions to $L_p$ now need to be counted for half of what they normally would since there are now twice as many (Table \ref{table:Diamond}).  Doing so yields an identical mathematical result from any other approach, but allows for full eight-fold speedup by only calculating one octant.  The exact weights for each basis atom in each volume of the algorithm are described in Table \ref{table:Diamond}.
\begin{table}[h]
	\begin{tabular}{ccccc}\hline
	
	%\tablewidth{0pt}
	\Tstrut Basis & Offset & Cube & Face & Axis\\
	Atom & (X,Y,Z) & Multiple & Multiple & Multiple\\\hline
	\Tstrut 1 & $(0,0,0)$ & 1 & 1 & 1\\[1.9pt]
	2 & $(\frac{1}{2},\frac{1}{2},0)$ & 1 & 1 & 2\\[1.9pt]
	3 & $(0,\frac{1}{2},\frac{1}{2})$ & 1 & 2 & 4\\[1.9pt]
	4 & $(\frac{1}{2},0,\frac{1}{2})$ & 1 & 2 & 2\\[1.9pt]
	5 & $(\frac{1}{4},\frac{1}{4},\frac{1}{4})$ & 0.5 & 1 & 2\\[1.9pt]
	6 & $(\frac{3}{4},\frac{3}{4},\frac{1}{4})$ & 0.5 & 1 & 2\\[1.9pt]
	7 & $(\frac{3}{4},\frac{1}{4},\frac{3}{4})$ & 0.5 & 1 & 2\\[1pt]
	8 & $(\frac{1}{4},\frac{3}{4},\frac{1}{4})$ & 0.5 & 1 & 2\\[1pt]
	9 & $(\frac{1}{4},\frac{1}{4},\frac{3}{4})$ & 0.5 & 1 & 2\\[1pt]
	10 & $(\frac{3}{4},\frac{3}{4},\frac{3}{4})$ & 0.5 & 1 & 2\\[1pt]
	11 & $(\frac{3}{4},\frac{1}{4},\frac{1}{4})$ & 0.5 & 1 & 2\\[1pt]
	12 & $(\frac{1}{4},\frac{3}{4},\frac{1}{4})$ & 0.5 & 1 & 2\Bstrut\\\hline
	%% Text for table notes should follow after the \enddata but before
	%% the \end{deluxetable}. Make sure there is at least one \tablenotemark
	%% in the table for each \tablenotetext.
	\end{tabular}
	\caption{Convential DIA lattice basis atoms (2-8) and additional tetrahedral basis atoms (9-12) for use in Algorithms \ref{alg:Sym} and \ref{alg:ParallelSym}\label{table:Diamond}.  The Cube Multiple represents the numerator used for the ($L_{cube/face/axis}$ += ) lines in psuedocode.}
\end{table}

\subsubsection{Extending the Exploitation of Symmetry: HCP}
\begin{figure}[ht]
	\caption{The HCP lattice as composed of four (red, blue, green, orange) inter-penetrating orthorhombic lattices.}
	\includegraphics[width=\linewidth]{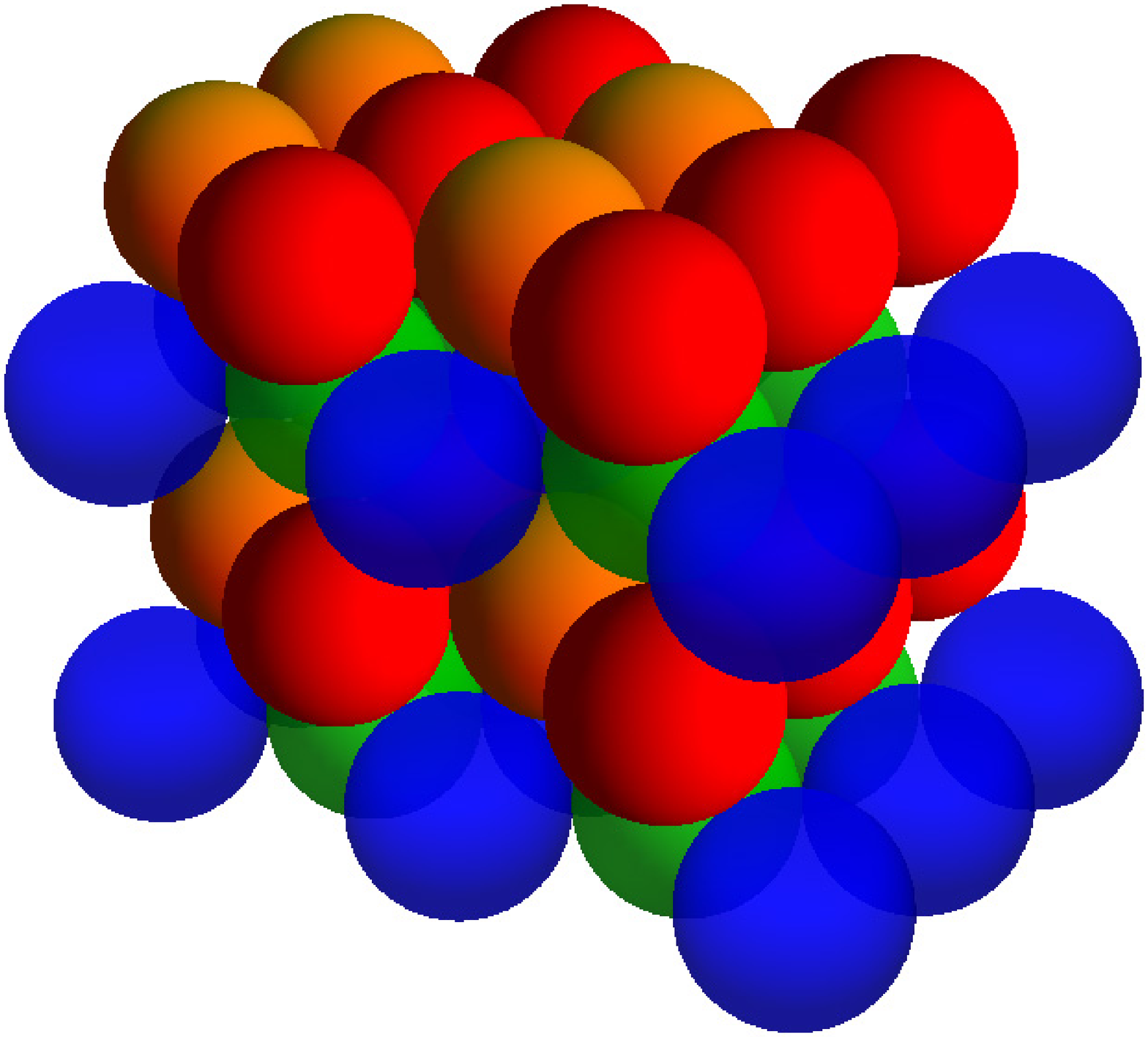}
	\label{im:HCPoff1}
\end{figure}

For HCP, due to the hexagonal nature of the lattice, a completely different approach is used. Using the fact that the HCP lattice has alternating layers (ABABAB) and those layers have alternating and repeating rows, the structure can be logically constructed as four inter-penetrating orthorhombic sub-lattices. One sub-lattice must be chosen to contain the $(0,0,0)$ position, whereas the others are identical in shape but offset from this first sub-lattice (Figure \ref{im:HCPoff1}).  An algorithm can be constructed to calculate these four sub-lattices separately, and each can be parallelized as before.

\begin{figure}[ht]
	\caption{Top-down view of the HCP lattice. The orange sub-lattice shown cuts unevenly across one of the yellow axes, where as the others either cut evenly or not at all.}
	\includegraphics[width=\linewidth]{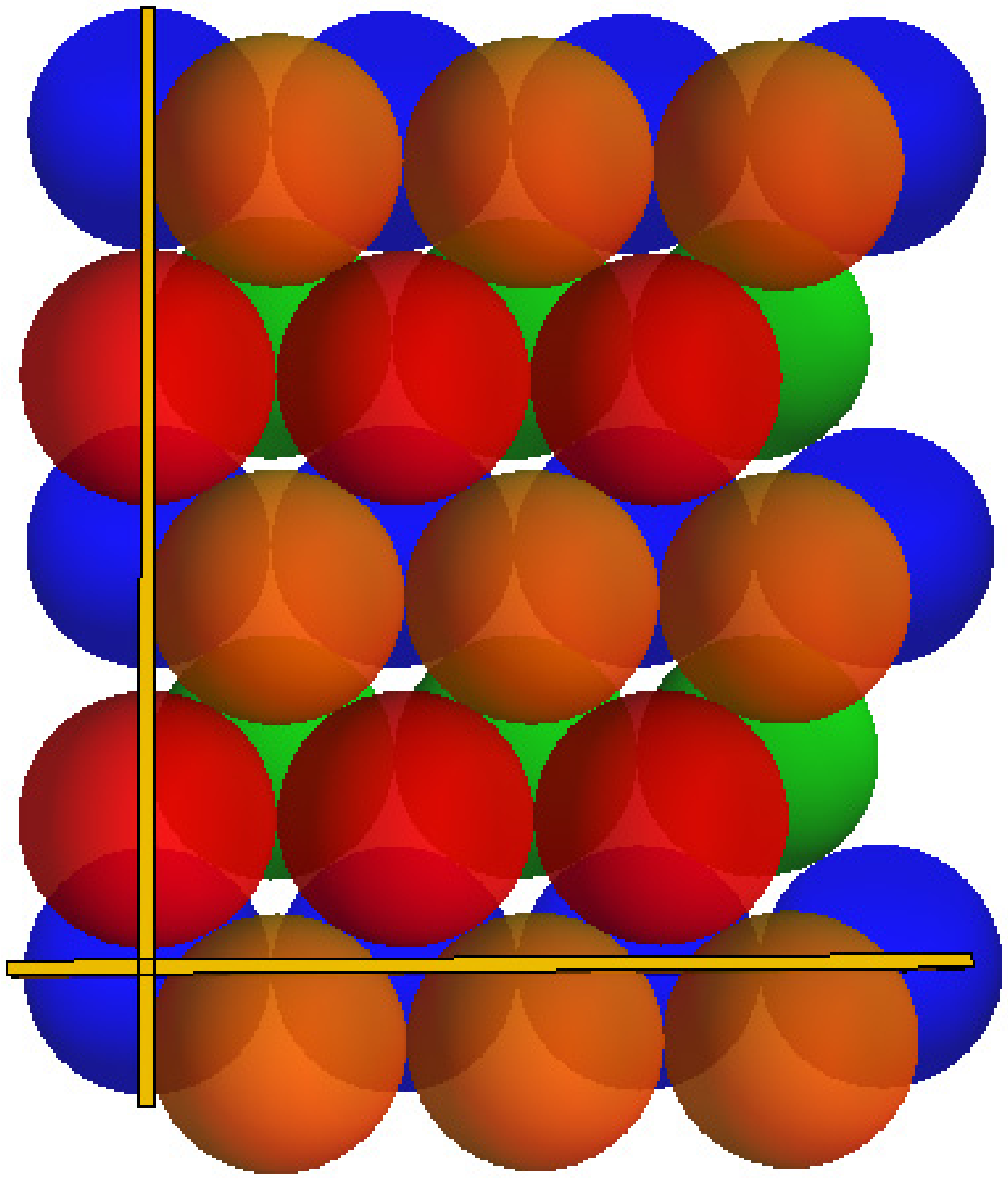}
	\label{im:HCPaxis}
\end{figure}

The symmetry of this lattice can be easily broken down into quartets, but using octants will present a similar challenge as the DIA lattice.  One of the four sub-lattices will extend slightly beyond what would be one of the faces between octants (Figure \ref{im:HCPaxis}), and distances to each atom from the origin are not the same across this axis.  The other three sub-lattices have atoms that either lie exactly on the faces, or completely within an octant.  The solution, similar to DIA, is to double the number of atoms in the only sub-lattice with unevenly shared atoms.  The positions of the extra atoms will be those that respect the rotational symmetry required for splitting the entire HCP lattice into equal octants.  As with DIA, the algorithm halves the value added to $L_p$ from each atom in this sub-lattice.

\subsubsection{Onionization}
\begin{figure}[ht]
	\caption{Left: Layers of computational ranges for different jobs shown stacking. The results are summed using the farthest first method.  Right: Each layer can be split into six volumes: three volumes for the cubic space (blue), two volumes for the face (red), and one volume for the axis (green).}
	\includegraphics[width=\linewidth]{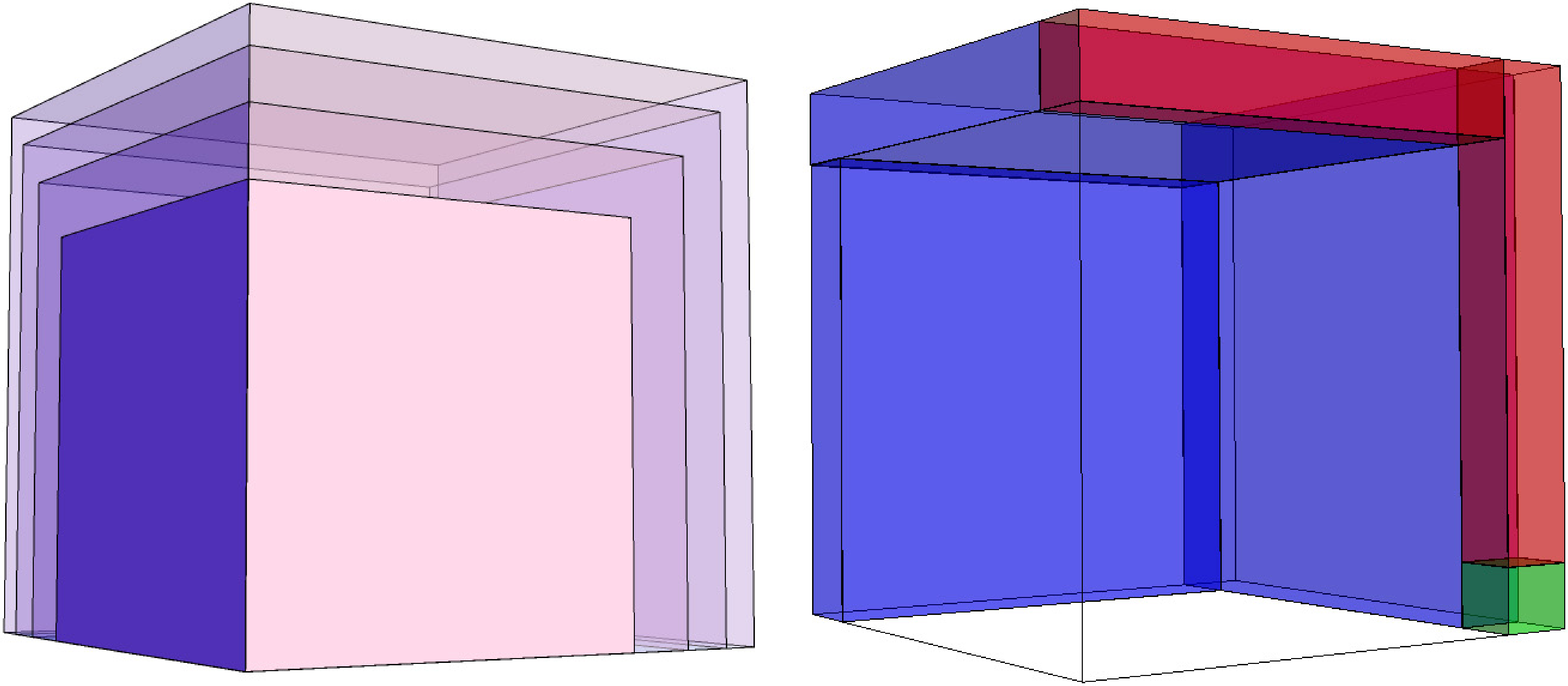}
	\label{im:onion}
\end{figure}

Running large, parallelized jobs on a cluster is convenient for solving large problems such as the algorithms described above.  However, software and hardware errors do occasionally occur which can result in many lost CPU hours.  As such, it is wise to break one large computation into many small ones.  The result is a series of jobs that stack like layers of a (cubic) onion that are gradually added to the problem set (Figure \ref{im:onion}, Left).  This has the added benefit of being able to reduce roundoff error for extremely small terms (i.e. those layers at greatest distance) if one performs the sum of each job's return value from smallest to greatest.

In the symmatrized version of the program, the expedient use of this method would involve breaking the onion layer down into six new volumes: three volumes that span the main cube volume from the inner layer to the new outer layer, two regions to cover the face, and one region to cover the axis (Figure \ref{im:onion}, Right).  This avoids having to check that the coordinates covered are outside of the previous layer, and the cost of entering and leaving the \textbf{for}--loops is negligible compared to the number of \textbf{if}--statements avoided.  Fortunately, the calculation of bases for each volume are identical to the non-onionized version.

\section{Results \& Applications} \label{sec:applications}

\subsection{Lennard-Jones Lattice Constants}

Using the symmatrized and parallelized algorithms described above, the Lennard-Jones lattice constants $L_p$ have been calculated in the SC, BCC, FCC, HCP, and DIA lattices (Table \ref{table:LJterms}). Terms with $p>9$ are computed to 32 decimal digits, convenient for quadruple precision calculations.  Those with $p\leq9$ are computed to lower precision due to computational limits (Figure \ref{im:conv}).  In addition to extending the precision of these constants, there are corrections to terms with $p<12$ previously published \citep{PhysRev.137.A152},\citep{PhysRevB.73.064112}.  The total speedup achieved going from the brute force method to the symmetric, parallel program was ${\sim}29$ fold (Table \ref{table:speedup}). 

To validate these results, a Mathematica program similar to Algorithm \ref{alg:BrokenDown} was used for several terms with $p\geq12$ using infinite precision in all five lattices.  Higher order terms were chosen because of the faster convergence of higher $p$ values, and the comparative slowness of Algorithm \ref{alg:BrokenDown} with the use of infinite precision.  The results all agreed to the given precision in Table \ref{table:LJterms}.

Moreover, a new method of computing any value that depends on distance between atoms in a crystal lattice has been created and optimized.  This same algorithm can be tailored to look at other crystal energy functions, such as the Buckingham potential, SW potential, and others.  

%\vspace{5mm}
\begin{table}[h]
	\begin{tabular}{ccc}\hline
		\Tstrut Algorithm & Fraction of $D^3$ & Effective Speedup\\
		 & terms in $L_p$ & per CPU core \\\hline
		\Tstrut Simple (\ref{alg:simple}) & $1$ & $1$\\[1.9pt]
		Broken Down (\ref{alg:BrokenDown}) & $1$ & $2.671$\\[1.9pt]
		Parallel (\ref{alg:Parallel}) & $1$ & $3.476$\\[1.9pt]
		Symmetric Parallel (\ref{alg:ParallelSym}) & $0.1249$ & $28.99$\Bstrut\\\hline
	\end{tabular}
	\caption{Speedup achieved for each algorithm in the SC lattice, as normalized to the calculation time of Algorithm \ref{alg:simple}.  Results are similar for any distance-dependent calculation. Gains in Algorithm \ref{alg:BrokenDown} are from avoiding the \textbf{if} statements. Likewise for Algorithm (\ref{alg:Parallel}) but most threads are also able to avoid jumping into and out of \textbf{for} loops, which also avoids calculating some of the same components of the distance value.  Algorithm \ref{alg:ParallelSym} combines the advantages of parallelism and 8-fold symmetry for the greatest gains.  All values of speedup are given per CPU core.\label{table:speedup}}
\end{table}

\subsection{Other Classical Potentials}

As the results in Table \ref{table:speedup} are normalized to 1, similar speed-up values should be attainable for applications of this approach to other crystal calculations.  For example, the SW potential is fit with as many as nine parameters:

\begin{equation*}
\begin{split}
U_{tot}&={\epsilon}\left[\sum_{i<j}f_2(r_{ij}/\sigma)+\sum_{\substack{i\neq{j}\\j<k}}f_3({r}_{ij}/\sigma,{r}_{ik}/\sigma,\theta_{ijk})\right]\\
f_2(r)&=\begin{cases}A\left(B{r^{-p}}-r^{-q}\right)exp\left[(r-a)^{-1}\right],r<a\\
0,r\geq{a}
\end{cases}\\
f_3({r}_{ij}&,{r}_{ik},\theta_{ijk})={\lambda}\left(\cos\theta_{ijk}+\frac{1}{3}\right)^2{\exp\left(\frac{\gamma}{r_{ij}-a}\right)}\\&\times\exp\left(\frac{\gamma}{r_{ik}-a}\right)
\label{eq:SW}
\end{split}
\end{equation*}

Fitting these parameters over many lattice sites and simulation requires lengthy computation, but it is also the case that the range of the potential is cut off at some arbitrary value (in this case, $a$).  Indeed the cutoff is typically so short that only nearest- or next-to-nearest neighbors contribute to the total energy.  Relaxing this parameter would allow simulation of more effects from vacancy or interstitial events.  The algorithms described above can be used to compensate for the additional calculations, resulting in a potentially more transferable fit.

\subsection{Applications to Crystal Defects}

To simulate defects, one cannot use an algorithm for calculating over lattice sites in a perfect crystal.  For example, to test or fit parameters for the creation energy of a Frenkel pair \cite{Frenkel1926}, the atom at the origin in the algorithms in this paper can be instead walked along (or integrated over) the path of defect creation.  A study is currently underway by the author to compare experimental data with recent parameterizations of the SW potential \cite{Pizza2013} and MD simulations of the threshold displacement energies for silicon \cite{PhysRevB.78.045202}.  This study relies on the algorithms presented in this paper to produce timely and accurate results.

In addition to point defects, plane defects can be simulated by displacing an entire algorithmic volume (as in Figures \ref{im:split1} or \ref{im:onion}) for as many planes as desirable. This allows for a faster way to test the transferability of plane defects to other parameterizations of potentials.  Either point or plane defects could be implemented as single occurrences, or uniform occurrences at regular intervals. Uniformly spread defects or point defects at the origin would still allow use of all the algorithms presented.

\section{Conclusions}

A series of algorithms has been developed for fast calculation of any distant-dependent property of lattices, including imperfect lattices.  The algorithms are adaptable to simulate point or plane defects for fitting or testing transferability of parameters in classical potential formulas, and the speedup achieved allows for relaxation of cut-off parameters. These algorithms can be used in serial or parallel, with the greatest speedup achievable through parallelization.  As an example of the power of the new algorithm, the Lennard-Jones lattice constants were determined up to 32 significant figures, extending their known precision, and in some cases correcting published figures.

\section{Acknowledgments}
The author would like to express his sincere gratitude to his advisor, Jodi Cooley, for her continuous support of his studies. The author would like to thank Randy Scalise and John Fattaruso for the inspiration and assistance with this project. The author is grateful to the SMU Center for Scientific Computation and Amit Kumar for the incredible amount of support and CPU time that made this all possible.  The author would also like to thank the SuperCDMS (Cryogenic Dark Matter Search) collaboration and its members for their feedback.

\bibliography{references}

\onecolumngrid
\begingroup
\squeezetable
\begin{sidewaystable} % <-- HERE
	%\centering
	\begin{tabular}{clllll}\hline
		\Tstrut$L_p$ & \multicolumn{1}{c}{SC} & \multicolumn{1}{c}{BCC} & \multicolumn{1}{c}{FCC} & \multicolumn{1}{c}{HCP} & \multicolumn{1}{c}{DIA} \\\hline
		\Tstrut $L_{4}$ & 16.53228 & 22.63872 & 25.33826 & 25.33908 & 10.23284 \\
		$L_{5}$ & 10.37752483 & 14.75850937 & 16.96751846 & 16.96843635 & 6.3127603582 \\
		$L_{6}$ & 8.40192397482754 & 12.2536678672923 & 14.4539210437445 & 14.4548972778416 & 5.11677158774719 \\
		$L_{7}$ & 7.4670577809188105309 & 11.054243479244464865 & 13.359387700742084043 & 13.360346776195552357 & 4.5944760255509476375 \\
		$L_{8}$ & 6.945807927226369624170778 & 10.35519790840251472712393 & 12.80193723137813255579318 & 12.80282185280989588716611 & 4.331913743971506684986912 \\
		$L_{9}$ & 6.6288591988867790990360972133 & 9.8945896563211153516496003879 & 12.492546702137558143156650385 & 12.493321725001781579567943092 & 4.1903721256503685465845227190 \\
		$L_{10}$ & 6.4261191025330890066321213261759 & 9.5644006153599478732928958387003 & 12.311245665477405791382158094686 & 12.311896233818981044642686360567 & 4.1110235994909590303207697180817 \\
		$L_{11}$ & 6.2922944992345673779692130757460 & 9.3132625373991001062237915286944 & 12.200920351277113166130939018073 & 12.201447099831954637516858217646 & 4.0654675989746082168420674657813 \\
		$L_{12}$ & 6.2021490450475185519304163922851 & 9.1141832680753588676564570885073 & 12.131880196544579708261946410532 & 12.132293769098917625885375250999 & 4.0389047128814160283254903749042 \\
		$L_{13}$ & 6.1405995800216921356289883683193 & 8.9518073185747151615181986692628 & 12.087726321352052662825461301813 & 12.088042550298439000808701615697 & 4.0232511870016901630637777123470 \\
		$L_{14}$ & 6.0981841257121521327529131655605 & 8.8167702284859198676408291902320 & 12.058991944350859312923039015626 & 12.059228255068241446619187406471 & 4.0139560884377807286889806009992 \\
		$L_{15}$ & 6.0687642950388921085943676325996 & 8.7029845599809255484889009743042 & 12.040024055099088629979906152137 & 12.040197144347223255169612620282 & 4.0084052364270646075922022605494 \\
		$L_{16}$ & 6.0482634695858416679464468171129 & 8.6062540475445294099631310987788 & 12.027354844018570329377774991884 & 12.027479419303856131335523169808 & 4.0050758707839297702973055370689 \\
		$L_{17}$ & 6.0339293163672074104867291112617 & 8.5235312504392982160783027553316 & 12.018809436710457796909425073283 & 12.018897719622859506995597162558 & 4.0030720422457274477625256930317 \\
		$L_{18}$ & 6.0238817078667147749258096087048 & 8.4525031686083817738457198340021 & 12.012998309665959588741240723167 & 12.013060023177408319000392309630 & 4.0018626537287036835660859738894 \\
		$L_{19}$ & 6.0168254563317377075012947930151 & 8.3913507914131177999359968485349 & 12.009019604439323572913310649568 & 12.009062224111209496811523499010 & 4.0011310801102128665086841693924 \\
		$L_{20}$ & 6.0118628308899457271005735996598 & 8.3386040056795629677517884205378 & 12.006280041326342657408789230664 & 12.006309158114658698075084825914 & 4.0006877092234562914662537575434 \\
		$L_{21}$ & 6.0083687575466831672374349304180 & 8.2930503704152943633396742334136 & 12.004384809362303299396537324646 & 12.004404510084773210744581189288 & 4.0004185815027828842036662413099 \\
		$L_{22}$ & 6.0059065261342911165963135002055 & 8.2536752180847796080069665089658 & 12.003068569322929886793306584316 & 12.003081784233296681711658420583 & 4.0002550042608394382922245458342 \\
		$L_{23}$ & 6.0041702400707480222133735785364 & 8.2196205348836491723537588485715 & 12.002151490974712110042308831699 & 12.002160286739322615229650383062 & 4.0001554703179378056666833678459 \\
		$L_{24}$ & 6.0029452081841294974019070764771 & 8.1901554754831630841615868276327 & 12.001510824939707072003031295877 & 12.001516638577045961535386165550 & 4.0000948484402457089307954562207 \\
		$L_{25}$ & 6.0020805203749133366722634113763 & 8.1646543519273306952128314265671 & 12.001062278709246141393049474196 & 12.001066097142027202635675470834 & 4.0000578966459919539177028099623 \\
		$L_{26}$ & 6.0014699724960860576163080284069 & 8.1425796159207988372706191903706 & 12.000747674897726915713658895972 & 12.000750168624485190911852889731 & 4.0000353574918921945136770397920 \\
		$L_{27}$ & 6.0010387522383048441317310970378 & 8.1234683158727891883161762963052 & 12.000526690212160028001079232466 & 12.000528310428505143624399658244 & 4.0000216015075983727843182332604 \\
		$L_{28}$ & 6.0007341210707894933633653237460 & 8.1069210710387173403772047642013 & 12.000371277553079701932705119123 & 12.000372325322411021726121333862 & 4.0000132018718538927954716341537 \\
		$L_{29}$ & 6.0005188792122114123860710817340 & 8.0925929383761217775918862369037 & 12.000261871447419639515536483048 & 12.000262546150133902013292069507 & 4.0000080707501767235974407653933 \\
		$L_{30}$ & 6.0003667748971840388354587433669 & 8.0801857499061731047494909917169 & 12.000184790059821196656550158876 & 12.000185222851788273771215661366 & 4.0000049351525392974917582258906 \\
		$L_{\infty}$ & 6 & 8 & 12 & 12 & 4 \Bstrut\\\hline
	\end{tabular}
	\caption{The Lennard-Jones lattice coefficients $L_p$ in the SC, BCC, FCC, HCP, and DIA lattices\label{table:LJterms}}
\end{sidewaystable} % <-- HERE
\endgroup
\twocolumngrid

\end{document}